\documentclass[preprint, aps]{revtex4}

\begin{document}
\draft
\title{Detection of local-moment formation using the resonant interaction between coupled quantum wires}
\author{V. I. Puller and L. G. Mourokh}
\address{Department of Physics and Engineering Physics \\
Stevens Institute of Technology, Hoboken, NJ 07030 }
\author{A. Shailos and J. P. Bird}
\address{Department of Electrical Engineering \\
\& Center for Solid State Electronics Research \\
Arizona State University, Tempe, AZ 85287-5706 }
\date{\today}

\begin{abstract}
{ We study the influence of many-body interactions on the transport characteristics of a novel device structure, consisting of a pair of quantum wires that are coupled to each other by means of a quantum dot. Under conditions where a local magnetic moment is formed in one of the wires, we show that tunnel coupling to the other gives rise to an associated peak in its density of states, which can be detected directly in a conductance measurement. Our theory is therefore able to account for the key observations in the recent study of T. Morimoto et al. [Appl. Phys. Lett. {\bf 82}, 3952 (2003)], and demonstrates that coupled quantum wires may be used as a system for the detection of local magnetic-moment formation.
 }
\end{abstract}

 \maketitle

The Kondo effect, involving the interaction between a localized
magnetic (spin) moment and free electrons, is one of the most
well-known manifestations of many-body behavior in solid-state
systems \cite{1}. Recently, interest in this effect has been
revived due to its importance for understanding the electrical
properties of mesoscopic quantum wires and dots \cite {2,3,4,5,6}.
In quantum dots, for example, the Coulomb blockade may be
exploited to confine an odd number of electrons on the dot, giving
rise to a net spin polarization that in turn plays the role of the
localized magnetic moment in the conventional Kondo effect
\cite{2}. In few electron dots, this effect can also be observed
for an even number of confined electrons, when the singlet and
triplet spin states are degenerate with each other \cite{4}. More
complicated behavior yet is found in non-equilibrium situations,
where tunneling involving higher spin states \cite{5} and a split
Kondo resonance \cite{6} have been observed.

Recently, there has been much interest generated by the suggestion
\cite{7,8,9} that the Kondo effect may also be responsible for the
so called ''0.7-structure'', observed near the conductance
threshold of quantum point contacts \cite{10,11}. The origin of
the 0.7 structure has served as the subject of intense debate for
more than a decade, since it cannot be accounted for within a
single-particle description. Recently, however, it was proposed
that this structure is associated with the formation of a
localized spin moment in the point contact, which develops as the
electron density in the channel is driven towards full depletion
\cite{8,9,b}. Independent support for this idea was suggested by
the results of a recent experimental study \cite{12}, in which
transport through the device shown schematically in the Inset to
Fig. 1 was investigated. (The black regions in this figure
represent sub-micron scale metal gates, which are deposited on the
surface of an ultra-high mobility GaAs/AlGaAs quantum well.) The
key observation in this experiment was obtained by applying fixed
voltages to gates 1 - 3 to form a point contact (which we refer to
hereafter as the ''fixed wire''), whose conductance was then
measured while varying the voltage ($V_{g}$) applied to gate 4
(which forms what we term the ''swept wire''). As the voltage
applied to the swept wire was varied over a wide range, little
noticeable influence was observed on the fixed-wire conductance,
as long as the swept wire supported at least one propagating mode.
Over the narrow range of voltage where the swept wire pinched off,
however, a resonant peak was observed in the conductance of the
fixed wire. The following characteristics of this resonant
interaction were noted by the authors of Ref. \cite{12}: (i) While
variation of the voltages applied to gates 1 - 3 could be used to
modify the pinch-off condition for the swept wire, the peak in the
conductance of the fixed wire was always found to remain
correlated to this pinch-off condition; (ii) In all cases, the
peak was manifest as an {\it enhancement} of the {\it conductance}
of the fixed wire; (iii) the amplitude of this enhancement was
roughly 0.1 $e^{2}/h$, independent of the conductance of the fixed
wire, which was varied from $\sim$ 1 - 12 $e^{2}/h$ in experiment;
(iv) The peak was no longer observed at temperatures above 4 K,
where the conductance quantization in the wires was also washed
out, and; (v) By grounding either gate 2 or 3, the same experiment
could be performed, although in this case the coupling between the
two wires was provided by a region of two-dimensional electron
gas, instead of a quantum dot. Even in this case, however, the
conductance of the fixed wire was found to show a peak as $V_{g}$
was varied, and the characteristics of this peak were similar to
those found when the coupling between the wires was provided by
the quantum dot.

In this Letter, we propose a theoretical explanation for the results of Ref.
\cite{12}, the key feature of which is a tunnel-induced correlation that arises
from the interaction between a localized magnetic moment in the swept wire and
conducting states in the fixed wire. An important feature of the device in Ref.
\cite{12} was that the elastic mean free path was more than an order of
magnitude larger than the inter-wire separation, so that electrons could be
exchanged between the two wires without significant impurity scattering. Starting from the assumption that a
localized magnetic moment is formed in the swept wire as it is driven near to pinch off \cite{7,8,9,b}, we
demonstrate that tunnel coupling of this moment to the fixed wire gives
rise to the appearance of a resonance in its density of states. The resonance is
in turn manifested as a peak in the conductance of the fixed wire and, based
upon this model, we can predict three-distinct regimes of behavior for the
coupled-wire system. (a) The swept wire has not yet reached pinch off and a
localized magnetic moment is not yet formed. In this case, the possibility of
charge transfer between the wires causes only additional broadening of the
states of the fixed wire. This broadening should be much smaller than that
caused by the coupling to the external circuit, however, and so should not
significantly affect the conductance of the fixed wire; (b) The gate voltage on
the swept wire is close to the pinch-off value, and a correlated many-body state
(localized magnetic moment) is formed in this wire. Further below, we
demonstrate that the formation of this spin moment is manifested as a sharp peak
in the density of states, and so in the conductance, of the fixed wire; (c) The
swept wire is completely pinched off and depopulated of electrons. Under such
conditions, tunneling of electrons between the two wires is strongly suppressed
and the density of states in the fixed wire is essentially the same as in case
(a) above (although there is now no additional broadening due to the exchange of
carriers between the wires).

The linear-response conductance of the fixed wire can be expressed in terms of
the density of states in this wire by means of the following generalization of
the Landauer conductance formula \cite{MeirWingreen}:
\begin{equation}
g=\frac{e^{2}}{h }\sum_{\sigma }\int d\epsilon \left( -f^{\prime
}(\epsilon )\right) \Gamma _{\sigma }\rho _{\sigma }(\epsilon ).
\label{conductance}
\end{equation}
Here, $f(\epsilon )$ is the Fermi distribution function,{\bf \ }$\Gamma_{\sigma
}$ describes the coupling between the wire and the leads that connect it to the
external circuit (for simplicity we assume uniform coupling to different lead
states), and $\rho _{\sigma }(\epsilon )$ is the density of states per spin in
the fixed wire that, in turn, can be expressed in terms of the retarded Green's
function in this wire, ${\bf G}^{r}$
\begin{equation}
\rho _{\sigma }(\epsilon )=-\frac{1}{\pi }%
\mathop{\rm Im}%
\left[ tr\left\{ {\bf G}^{r}(\epsilon )\right\} \right] .  \label{density}
\end{equation}

To calculate this Green's function, we employ a procedure similar to that
of \ Refs. \cite{Anderson,Lee} with the Hamiltonian given by
\begin{eqnarray}
\widehat{H} &=&\sum_{\sigma }\epsilon _{\sigma }n_{\sigma
}+\frac{1}{2}%
U\sum_{\sigma }n_{\sigma }n_{\overline{\sigma }}+\sum_{q,\sigma }\epsilon
_{q\sigma }n_{q\sigma }+\sum_{k,\sigma }E_{k\sigma }c_{k\sigma
}^{+}c_{k\sigma }+  \label{Hamiltonian} \\
&&+\sum_{k,\sigma }\left( V_{k\sigma }c_{k\sigma }^{+}a_{\sigma
}+V_{k\sigma }^{\ast }a_{\sigma }^{+}c_{k\sigma }\right)
+\sum_{k,q,\sigma }\left( v_{kq\sigma }c_{k\sigma }^{+}a_{q\sigma
}+v_{kq\sigma }^{\ast }a_{q\sigma }^{+}c_{k\sigma }\right)
\nonumber
\end{eqnarray}
The first two terms in the Hamiltonian, Eq.(\ref{Hamiltonian}), are the regular
Anderson Hamiltonian \cite{Anderson} describing the localized magnetic moment
which is formed in the swept wire, when it is biased close to its pinch-off
point. ( Here, $\epsilon _{\sigma }$ is the energy of electron having spin
$\sigma $, $\overline{\sigma }$ is the quantum spin state opposite to $\sigma $, $U$ is the Coulomb energy, and $n_{\sigma }=a_{\sigma }^{+}a_{\sigma
}$, where \ $a_{\sigma }^{+}$ and $a_{\sigma }$ are electron creation and
annihilation operators in the swept wire.) In general, when the wire is not
pinched off, there is a continuum of states that is characterized by different
momenta along the wire. However, in the presence of electron-electron
interactions, and under conditions leading to the formation of the localized
magnetic moment, only one of these states is occupied.

The third term of the Hamiltonian describes free electrons in the fixed wire,
which, besides their spin quantum number, are also characterized by their
longitudinal momentum along the wire, $q$. $a_{q\sigma }^{+}$ and $a_{q\sigma
}$ are creation and annihilation operators for an electron with momentum $q$
along the wire and spin $\sigma $, having the energy $\epsilon _{q\sigma }$,
while $n_{q\sigma }=a_{q\sigma }^{+}a_{q\sigma }$. Since this wire is typically not biased close to pinch off, it may contain electrons in states characterized
by different values of $q$.

The fourth term in the Hamiltonian describes electrons in the quantum dot that
mediates the interaction between the wires, characterized by a set of quantum
numbers $k$. The corresponding creation and annihilation operators are
$c_{k\sigma }^{+}$ and $c_{k\sigma }$, respectively, and the eigenenergies are
$E_{k\sigma }$. We emphasize, however, that, because of the very general form of
this term, no significant modification of the Hamiltonian, Eq.
(\ref{Hamiltonian}), is necessary to
 describe the situation when gates 2 and 3 in Figure 1 are grounded.

The two last terms in the Hamiltonian describe tunnel coupling between the swept
wire and quantum dot (with matrix element $ V_{k\sigma }$), and the fixed wire and quantum dot (with matrix element $ v_{kq\sigma }$), respectively.

At this point, we introduce the following retarded Green's functions
(superscript ''$r$''
is omitted hereafter):\newline
The Green's function of electrons in the fixed wire
\begin{equation}
G_{qq_{1}\sigma }(t)=-i\theta (t)\left\langle \left[ a_{q\sigma
}(t),a_{q_{1}\sigma }^{+}(0)\right] _{+}\right\rangle ;  \label{Green}
\end{equation}
The mixed Green's function describing tunneling between the fixed wire and the
quantum dot
\begin{equation}
g_{kq_{1}\sigma }(t)=-i\theta (t)\left\langle \left[ c_{k\sigma
}(t),a_{q_{1}\sigma }^{+}(0)\right] _{+}\right\rangle ;
\end{equation}
The mixed Green's function for electrons tunneling between the swept wire and the fixed wire (via the quantum dot)
\begin{equation}
G_{\sigma ;q_{1}\sigma }(t)=-i\theta (t)\left\langle \left[ a_{\sigma
}(t),a_{q_{1}\sigma }^{+}(0)\right] _{+}\right\rangle ;
\end{equation}
The two-particle Green's function, factorized in the Hartree approximation
\begin{equation}
G_{\overline{\sigma }\sigma ;q_{1}\sigma }(t)=-i\theta (t)\left\langle \left[
n_{\overline{\sigma }}(t)a_{\sigma }(t),a_{q_{1}\sigma }^{+}(0)\right]
_{+}\right\rangle =\left\langle n_{\overline{\sigma }}\right\rangle
G_{\sigma ;q_{1}\sigma }(t),
\end{equation}
where $\theta (t)$ is the unit step function. In addition to these definitions,
calculation of the population of the swept wire, $\left\langle n_{\sigma
}\right\rangle $, also requires knowledge of the electron Green's function in
the swept wire, as has been discussed in \cite{Anderson}.

The Fourier transformed equations of motion for the above Green's functions
are given by\qquad\
\begin{eqnarray}
\left( \epsilon -\epsilon _{q\sigma }\right) G_{qq_{1}\sigma }(\epsilon )
&=&\delta _{q,q_{1}}+\sum_{k}v_{kq\sigma }^{\ast }g_{kq_{1}\sigma
}(\epsilon
), \\
\left( \epsilon -E_{k\sigma }\right) g_{kq_{1}\sigma }(\epsilon )
&=&V_{k\sigma }G_{\sigma ;q_{1}\sigma }(\epsilon
)+\sum_{q_{2}}v_{kq_{2}\sigma }G_{q_{2}q_{1}\sigma }(\epsilon ),  \nonumber
\end{eqnarray}
and
\[
\left( \epsilon -\epsilon _{\sigma }-U\left\langle n_{\overline{\sigma }%
}\right\rangle \right) G_{\sigma ;q_{1}\sigma }(\epsilon
)=\sum_{k}V_{k\sigma }^{\ast }g_{kq_{1}\sigma }(\epsilon ).
\]
This set of equations allows us to obtain the diagonal elements of the
Green's function in the fixed wire as
\begin{equation}
G_{qq\sigma }(\epsilon )=\frac{1}{\epsilon -\epsilon _{q\sigma }+i\Delta
_{q\sigma }}+\frac{\left| T_{q\sigma }(\epsilon )\right| ^{2}}{\left(
\epsilon -\epsilon _{q\sigma }+i\Delta _{q\sigma }\right) ^{2}\left(
\epsilon -\epsilon _{\sigma }-U\left\langle n_{\overline{\sigma }%
}\right\rangle +i\left( \Delta _{\sigma }+\pi _{\sigma }\right) \right) },
\label{Result1}
\end{equation}
where the linewidths $\Delta _{q\sigma }$, $\Delta _{\sigma }$, $\pi
_{\sigma }$ are the imaginary parts of the corresponding self-energies, (we
neglect the level shifts \ due to their real parts) defined as
\begin{equation}
\Delta _{\sigma }=-%
\mathop{\rm Im}%
\left[ \sum_{k}\frac{\left| V_{k\sigma }\right| ^{2}}{\epsilon -E_{k\sigma }}%
\right] =\pi \sum_{k}\left| V_{k\sigma }\right| ^{2}\delta (\epsilon
-E_{k\sigma }),
\end{equation}
\begin{equation}
\delta _{q,q_{1}}\Delta _{q\sigma }=-%
\mathop{\rm Im}%
\left[ \sum_{k}\frac{v_{kq\sigma }^{\ast }v_{kq_{1}\sigma }}{\epsilon
-E_{k\sigma }}\right] =\delta _{q,q_{1}}\pi \sum_{k}\left| v_{k\sigma
}\right| ^{2}\delta (\epsilon -E_{k\sigma }),  \label{fixed}
\end{equation}
and
\begin{equation}
\pi _{\sigma }=-%
\mathop{\rm Im}%
\left[ \sum_{q}\frac{\left| T_{q\sigma }(\epsilon )\right| ^{2}}{\epsilon
-\epsilon _{q\sigma }-\Sigma _{q\sigma }(\epsilon )}\right] =\sum_{q}\frac{%
\left| T_{q\sigma }(\epsilon )\right| ^{2}\Delta _{q\sigma }}{\left[
\epsilon -\epsilon _{q\sigma }\right] ^{2}-\left[ \Delta _{q\sigma }\right]
^{2}}.
\end{equation}
Here, $T_{q\sigma }(\epsilon )$ is the matrix element describing electron
transfer from one of the wires into the quantum dot and, subsequently, to the
other wire and is given by
\begin{equation}
T_{q\sigma }(\epsilon )=\sum_{k}\frac{v_{kq\sigma }^{\ast }V_{k\sigma }}{%
\epsilon -E_{k\sigma }}.  \label{matrix element}
\end{equation}
The Kronecker $\delta _{q,q_{1}}$ appears in Eq. (\ref{fixed}) since we neglect
mixing of the fixed wire states when the electron transfers to the quantum dot
and back. This condition is easily satisfied by appropriate choice of the
eigenstates in the fixed wire (see similar discussion in Ref. \cite{Kondo}).

There are two features of Eq. (\ref{Result1}) that distinguish it physically
from the result of Anderson, Ref. \cite{Anderson}, for the polarization of the
free-electron bands in the proximity of a localized magnetic moment (which is
formally the same as the problem we consider). Firstly, Eq. (\ref{Result1})
reflects changes in the states of the fixed wire only, instead of all regions in
proximity to the localized moment. The influence of these other regions is
accounted for by incorporating the linewidths $\Delta_{q\sigma }$, $\Delta
_{\sigma }$, $\pi _{\sigma }$, which already account for the quantum dot region.
Secondly, the matrix element $T_{q\sigma}(\epsilon )$ does not describe the
hybridization leading to the formation of the localized magnetic moment, but
refers instead to ballistic tunneling of electrons between the two quantum
wires, via Eq. (\ref{matrix element}).

Using Eq. (\ref{Result1}) we can obtain the density of states in the fixed
wire, Eq. (\ref{density}), as
\begin{equation}
\rho _{\sigma }(\epsilon )=-\frac{1}{\pi }
\mathop{\rm Im}
\int dqG_{qq\sigma }(\epsilon )=-\frac{1}{\pi }
\mathop{\rm Im}
\int d\epsilon _{q\sigma }\rho _{\sigma }^{0}(\epsilon _{q\sigma
})G_{qq\sigma }(\epsilon ),  \label{density1}
\end{equation}
with $\rho _{\sigma }^{0}(\epsilon _{q\sigma })$ being the density of states in
the fixed wire, in the absence of any coupling to its leads or to the dot.
Neglecting the weak momentum dependence of the tunneling matrix element, and the
linewidths, we can carry out the integration in Eq. (\ref
{density1}) to obtain
\begin{eqnarray}
\rho _{\sigma }(\epsilon ) &=&\overline{\rho }_{\sigma }(\epsilon
)-\frac{d
\overline{\rho }_{\sigma }(\epsilon )}{d\epsilon }\frac{\left| T\right|
^{2}\left( \epsilon -\epsilon _{\sigma }-U\left\langle n_{\overline{\sigma }%
}\right\rangle \right) }{\left( \epsilon -\epsilon _{\sigma }-U\left\langle
n_{\overline{\sigma }}\right\rangle \right) ^{2}+\left( \Delta _{\sigma
}+\pi _{\sigma }\right) ^{2}}-  \nonumber \\
&&-\left[ \frac{1}{\pi }\frac{d}{d\epsilon }\int d\epsilon _{q\sigma
}\frac{
\rho _{\sigma }^{0}(\epsilon _{q\sigma })\left( \epsilon -\epsilon _{q\sigma
}\right) }{\left( \epsilon -\epsilon _{q\sigma }\right) ^{2}+\Delta
_{q\sigma }^{2}}\right] \frac{\left| T\right| ^{2}\left( \Delta _{\sigma
}+\pi _{\sigma }\right) }{\left( \epsilon -\epsilon _{\sigma }-U\left\langle
n_{\overline{\sigma }}\right\rangle \right) ^{2}+\left( \Delta _{\sigma
}+\pi _{\sigma }\right) ^{2}},  \label{density2}
\end{eqnarray}
where
\begin{equation}
\overline{\rho }_{\sigma }(\epsilon )=-\frac{1}{\pi }
\mathop{\rm Im}
\int d\epsilon _{q\sigma }\rho _{\sigma }^{0}(\epsilon _{q\sigma })\frac{1}{%
\epsilon -\epsilon _{q\sigma }+i\Delta _{q\sigma }}=\frac{1}{\pi }\int
d\epsilon _{q\sigma }\frac{\rho _{\sigma }^{0}(\epsilon _{q\sigma })\Delta
_{q\sigma }}{\left( \epsilon -\epsilon _{q\sigma }\right) ^{2}+\Delta
_{q\sigma }^{2}}
\end{equation}
is the density of states in the fixed quantum wire when the wires are uncoupled.
The last term on the right side of Eq. (\ref{density2}) is an order of magnitude
smaller than the one that precedes it, and can, therefore, be omitted. The most
essential feature of Eqs. (\ref{Result1}) and (\ref{density2}) is the presence
of an additional resonance in the density of states of the fixed wire, which
appears at the energy $\epsilon _{eff}=\epsilon _{\sigma }+U\left\langle
n_{\overline{\sigma }}\right\rangle $, which is the same as the energy of the
resonant state formed in the swept wire. This resonance therefore only appears
when a localized magnetic moment is formed in the swept wire, i.e. when
$\left\langle n_{\sigma }\right\rangle \simeq 1$ for just one state of the
swept wire, while all its other states are depopulated and so do not contribute
to the density of states, Eq. (\ref{density2}). An analysis of Eq.
(\ref{density2}) allows us to account for the main observations in the
experiment in Ref. \cite{12} (refer to observations (i) - (v) in the
introduction to this Letter): (i) the resonance occurs only as the swept wire
is pinched off, i.e. when a localized magnetic moment is formed in this
wire, leading to a sharp peak in its density of states that is located close to
the Fermi level. (ii) The essential feature of Eq. (\ref{density2}) is that it
contains the derivative of the density of states in the fixed {\it wire}. This
{\it one-dimensional} density of states depends on energy as $\overline{\rho
}_{\sigma }(\epsilon )\sim\rho_{\sigma }^{0}(\epsilon )\sim 1/\sqrt{\epsilon }$
and, accordingly, its derivative is negative. Consequently, the second term of
Eq. (\ref{density2}) is positive and so gives rise to an {\it enhancement} of
the fixed-wire conductance (see Eq. (1)), in agreement with the observations in Ref.
\cite{12}. In this sense, we see that the conductance of the fixed wire serves
as a detector of the localized magnetic moment in the swept wire. (iii) The
amplitude of the additional term in Eq. (\ref{density2}) is proportional to the
tunneling probability between the two wires, $\left| T\right| ^{2}$, and should
therefore be predominantly limited by the height of the tunnel barrier that
forms in the swept wire. It should not depend significantly, however, on the
conductance of the fixed wire, which is also consistent with experiment. (iv)
Since the appearance of the additional conductance peak is the manifestation of
a many-body state, related to the ''0.7'' anomaly, it should disappear at higher
temperatures where this feature is no longer observed. This is consistent with
the temperature-dependence of the conductance resonance reported in Ref.
\cite{12}. (v) We emphasize again that our analysis does not involve any
assumptions about the energy level structure in the quantum dot region. In
particular, our formalism should also be valid in the case where the region
between two wires is comprised of two-dimensional electron gas.

For further clarification, we have performed numerical
calculations of the correction to the conductance of the fixed
wire, resulting from the formation of the resonant energy level in
the swept wire. Substituting the first two terms of Eq. (15) into
Eq. (1), we obtain the conductance of the fixed wire in the form

\begin{equation}
g = \bar{g} + \Delta g,
\end{equation}

where the correction to the conductance, $\Delta g$, is given by

\begin{equation}
\Delta g = {2e^2\over\hbar}{|T|^2\over 4}\sum_{n=0}^{N-1}{1\over
E_F-\epsilon_0-n\hbar\omega_y}\sum_{\sigma}{E_F-\epsilon_{\sigma}-U\langle
n_{\bar{\sigma}}\rangle\over (E_F-\epsilon_{\sigma}-U\langle
n_{\bar{\sigma}}\rangle)^2+(\Delta_{\sigma}+\pi_{\sigma})^2}.
\end{equation}

Here, $E_F$ is the Fermi energy, $\hbar\omega_y$ is the energy of
transverse confinement in the fixed wire, $\epsilon_0$ is the
energy of its lowest subband, and $N$ is the number of occupied
subbands. The total linewidth $\Gamma = \Delta_{\sigma}
+\pi_{\sigma}$ is assumed to be $\sigma$-independent. The
numerically calculated value of $\Delta g$ is shown in Fig. 1, as
a function of the separation between the Fermi energy and the
energy of the resonant state in the swept wire. In order to obtain
this figure, we used the following set of parameters: $U = 0.6
meV, E_F-\epsilon_0 = 10 meV$, and $\hbar\omega_y = 3 meV$
\cite{8,9}. The total linewidth and transmission coefficient are
taken to be $\Gamma = 0.02 meV$ and $|T| = 0.08 meV$,
respectively. The former value is chosen to be much less than the
Coulomb energy, $U$, which is necessary for the local magnetic
moment to form \cite{8,9}. In fact, the value $\Gamma = 0.02 meV$
is commonly used in the estimation of the level broadening in the
quantum dot systems. One can see from Fig. 1 that the calculated
correction to the conductance peaks at a positive value when the
separation between the resonant state and the Fermi energy is
roughly equal to $\Gamma$, and that the height of the peak is
about 0.06 $2e^2/h$. The correction becomes negative as the energy
of the resonant state passes through the Fermi energy and this
resonant state disappears thereafter. Our calculations reproduce
the qualitative character of the experimental peak in Ref.
\cite{12}, such as its absolute magnitude, the relatively slow
growth on the right side of the curve, and the sharp drop on its
left side. Quantitative agreement is reached with the reasonable
set of parameters.

In conclusion, we have studied the influence of many-body
interactions on the conductance of the coupled-quantum-wire system
investigated in Ref. \cite{12}. Our model considers the influence
of tunnel coupling between the wires, under conditions where a
local moment is formed in one of them by biasing it close to its
pinch-off condition. The tunnel coupling is shown to give rise to
an associated peak in the density of states of the fixed wire,
which is manifested in turn as a peak in its conductance. Our
numerical calculations reproduce the qualitative character of the
experimental peak and give quantitative agreement with experiment
for reasonable choices of the model parameters. Our simple theory
is therefore able to account for the key observations of Ref.
\cite{12} and demonstrates that the system of coupled quantum
wires investigated here may serve as a detector of local-moment
formation.

\newpage
\begin{center}
Figure Caption \\
\end{center}

Figure 1. Correction to conductance of the fixed wire as a
function of the separation between the Fermi energy and the energy
of the resonant state formed in the swept wire. Inset: Schematic
illustration of the split-gate device studied in Ref. \cite{12}.
Black regions are metal gates that are deposited on the surface of
an ultra-high mobility quantum well and which are used to form a
pair of quantum point contacts that are coupled by a quantum dot.
For further details on the device structure, and its
characterization, we refer the reader to Ref. \cite{12}.

\newpage
\begin{center}
References \\
\end{center}

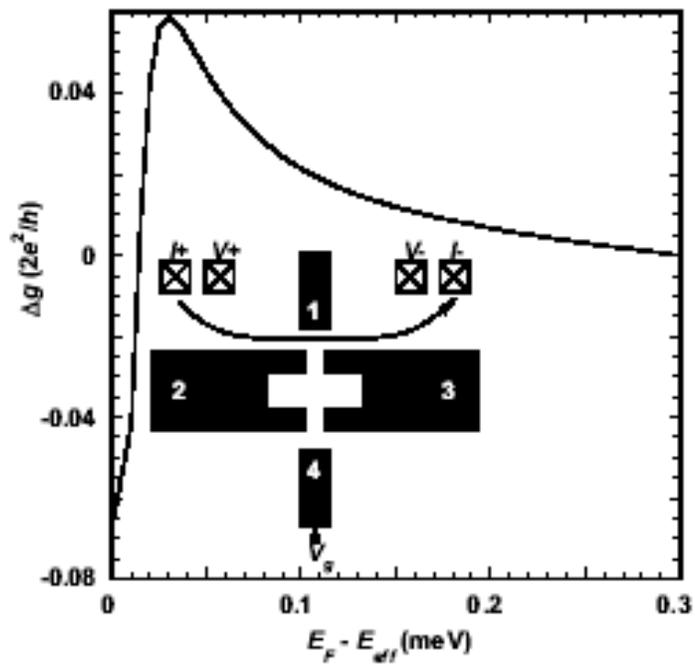
\end{document}